\documentclass[twocolumn]{aastex631}
\usepackage{amsmath, booktabs}
\usepackage{rotating}
\begin{document}


\defcitealias{nico19}{GC19}

\title{Hypervelocity Stars Trace a Supermassive Black Hole in the Large Magellanic Cloud}

\author{Jiwon Jesse Han}
\affiliation{Harvard-Smithsonian Center for Astrophysics\\
60 Garden St , Cambridge MA \\
}

\author{Kareem El-Badry}
\affiliation{Department of Astronomy, California Institute of Technology\\
1200 East California Blvd, Pasadena, CA}

\author{Scott Lucchini}
\affiliation{Harvard-Smithsonian Center for Astrophysics\\
60 Garden St , Cambridge MA \\
}

\author{Lars Hernquist}
\affiliation{Harvard-Smithsonian Center for Astrophysics\\
60 Garden St , Cambridge MA \\
}

\author{Warren Brown}
\affiliation{Harvard-Smithsonian Center for Astrophysics\\
60 Garden St , Cambridge MA \\
}

\author{Nico Garavito-Camargo}
\affiliation{Center for Computational Astrophysics, Flatiron Institute, Simons Foundation\\
162 Fifth Avenue, New York NY\\
}

\author{Charlie Conroy}
\affiliation{Harvard-Smithsonian Center for Astrophysics\\
60 Garden St , Cambridge MA \\
}

\author{Re'em Sari}
\affiliation{Racah Institute for Physics, The Hebrew University, Jerusalem 91904, Israel\\
}

\begin{abstract}
Hypervelocity stars (HVSs) are produced by the Hills mechanism when a stellar binary is disrupted by a supermassive black hole (SMBH). The HVS Survey detected 21 unbound B-type main-sequence stars in the Milky Way's outer halo that are consistent with ejection via the Hills mechanism. We revisit the trajectories of these stars in light of proper motions from {\it Gaia} DR3 and modern constraints on the Milky Way -- Large Magellanic Cloud (LMC) orbit. We find that half of the unbound HVSs discovered by the HVS Survey trace back not the Galactic Center, but to the LMC. Motivated by this finding, we construct a forward-model for HVSs ejected from an SMBH in the LMC and observed through the selection function of the HVS Survey. The predicted spatial and kinematic distributions of simulated HVSs are remarkably similar to the observed distributions.  In particular, we reproduce the conspicuous angular clustering of HVSs around the constellation Leo. This clustering occurs because HVSs from the LMC are boosted by $\sim300\,{\rm km\,s^{-1}}$ by the orbital motion of the LMC, and stars launched parallel to this motion are preferentially selected as HVS candidates. We find that the birth rate and clustering of LMC HVSs cannot be explained by supernova runaways or dynamical ejection scenarios not involving a SMBH. From the ejection velocities and relative number of Magellanic vs. Galactic HVSs, we constrain the mass of the LMC SMBH to be $10^{5.8^{+0.2}_{-0.4}}  M_{\odot}$ ($\simeq 6\times10^5 M_{\odot}$).
\end{abstract}

\section{Introduction} \label{sec:intro}

The Galactic halo contains a small number of stars that are traveling faster than the local escape velocity on trajectories that will carry them into intergalactic space. One mechanism for producing such hypervelocity stars (HVSs) is the \citet{Hills1988} mechanism: when a close stellar binary strays near a supermassive black hole (SMBH), one star can be captured, while the other is ejected at velocities that can reach $>1,000\,{\rm km\,s^{-1}}$. The captured star can produce an observable tidal disruption event or a variety of related classes of nuclear transients \citep[e.g.][]{Sari2010,bromley12, Alexander2017, Linial2023, Lu2023}. The ejected star is launched into the Galactic halo, where it can be observed for hundreds of Myr \citep[e.g.][]{Brown2015}.

The most successful observational search for HVSs to date was the HVS Survey \citep{Brown2006}. Following the serendipitous discovery of an unbound B star in the outer halo by \citet{Brown2005}, the HVS Survey carried out a systematic search for unbound B stars in the halo via spectroscopic follow-up of photometrically selected candidates. Because there has been no recent star formation in the halo, any young stars found in the halo must have traveled there from elsewhere. The HVS Survey targeted B stars because they are primarily young and can be efficiently selected photometrically. Over the course of almost a decade, the survey obtained spectra of more than 1400 blue halo sources, eventually leading to the discovery of 21 main-sequence B stars suspected to be unbound \citep{Brown2014}. These stars are at distances of $50-120$\,kpc and have masses of $2.5-4\,M_{\odot}$. The survey was 99\% complete within the imaging footprint of the Sloan Digital Sky Survey \citep[SDSS DR8;][]{sloan8} and had a well-defined selection function, making the sample well-suited for population modeling.

There are other processes, besides the Hills mechanism, that can accelerate stars to high velocities. The most important is  the \citet{Blaauw1961} kick, wherein a star is ejected from a binary when its companion explodes. For compact stars, such as white dwarfs and hot subdwarfs, this can produce ejection velocities comparable to the Hills mechanism \citep[e.g.][]{Justham2009, Shen2018, El-Badry2023}. Indeed, the second candidate HVS discovered was US 708 \citep[also known as HVS 2;][]{Hirsch2005}, which is a helium-burning hot subdwarf that was almost certainly ejected from a thermonuclear supernova \citep{Geier2015}. However, main-sequence B stars cannot be ejected from supernovae with such high velocities: their maximum SN ejection velocity is $\sim 500\,{\rm km\,s^{-1}}$, and the vast majority are ejected at much slower velocities of only a few tens of ${\rm km\,s^{-1}}$ \citep[e.g.][]{Bromley2009, Tauris2015, Renzo2019, Evans2020}. 3- and 4-body interactions in star clusters have been proposed as another mechanism to produce high-velocity stars \citep[e.g.][]{Leonard1991, Perets2012, Cabrera2023}, but these generally produce slower velocities than the Hills mechanism, and the predicted ejection rate of stars with velocities $>500\,{\rm km\,s^{-1}}$ is much lower than the observed HVS birth rate \citep{Brown2015}. 


One feature of the HVS sample discovered by the HVS Survey that has proved difficult to explain is the stars' anisotropic distribution on the sky: about half of the unbound HVSs are found around the Leo constellation, with 52\% (11/21) of the stars clustered within only 5\% of the HVS Survey footprint \citep[e.g.][]{Brown2009, Brown2012, Brown2014}. We refer to this clustering as the ``Leo Overdensity'' throughout this paper. A variety of models have been proposed to explain the Leo Overdensity, including an anisotropic Galactic gravitational potential \citep[e.g.][]{Kenyon2008}, binary SMBH in-spiral \citep{sesana06,sesana08}, imprints of the Milky Way's nuclear stellar disk \citep{Lu2010, Zubovas2013, Hamers17}, formation of HVSs in AGN outflows \citep{Wang2018}, or ejection of stars from a tidally disrupting dwarf galaxy \citep{Abadi2009, Piffl2011}. A particularly compelling idea was put forth by \citet{Boubert2016}, who studied the distribution of HVS from a hypothetical $1.7\times10^5M_{\odot}$ black hole in the LMC. They find that the resulting distribution is dipolar on the sky, due to the orbital motion of the LMC. In this work, we explore in detail the scenario wherein a significant fraction of HVSs come from the LMC.

Given their typical distance of $\sim 70$ kpc, HVSs have small proper motions of order $1\,\rm mas\,yr^{-1}$. This makes it challenging to trace HVSs back to their launching site. Leveraging proper motions from a multi-cycle {\it Hubble Space Telescope} program, \citet{Brown2015_pm} calculated orbits for the unbound stars discovered by the HVS Survey. They found trajectories consistent with a Galactic Center origin for most sources, but with large enough uncertainties to accommodate other launching sites. Comparably precise proper motions have now been measured by {\it Gaia} \citep{GaiaCollaboration2016, GaiaCollaboration2021}, with better-understood uncertainties and systematics. These measurements have allowed several authors to revisit the trajectories of HVSs \citep{Brown2018, Irrgang2018, Hattori2019, Kreuzer2020, Verberne2024}, with early results suggesting that several HVSs do {\it not} trace back to the Galactic Center. The most common interpretation has been that these stars are binary runaways ejected from the Galactic disk. While this hypothesis is difficult to falsify for any individual HVS, in several cases the inferred trajectories imply improbable launching sites in the outer disk, well beyond the solar circle.

A possible clue to the provenance of the wayward HVSs is offered by the star HE 0437-5439, also known as HVS 3 \citep{Edelmann2005}. This object is a $9\,M_{\odot}$ main-sequence star that is $\sim 60$ kpc from the Galactic Center, but only $\sim 15$ kpc from the Large Magellanic Cloud (LMC). It is unlikely to originate in the Milky Way, because this would require a flight time significantly longer than the lifetime of a $9\,M_{\odot}$ star. Consistent with early suggestions \citep{Gualandris2007, Bonanos2008, Przybilla2008}, {\it Gaia} proper motions now allow the star to be definitively traced back to the LMC \citep{Erkal2019}. Here, we explore whether the objects discovered by the HVS Survey (which, unlike HE 0437-5439, are all in the northern hemisphere) could similarly have been ejected from the LMC. 
Although this possibility has been considered previously, earlier studies were constrained by a lack of high-quality proper motions and/or inaccurate models of the Milky Way–LMC orbit. In contrast to these earlier works, we find that approximately half of the unbound stars identified by the HVS Survey are consistent with originating from the LMC. Therefore, we investigate the broader implications of a substantial fraction of the HVS sample being born in the LMC.

The remainder of this paper is organized as follows. In Section \ref{sec:methods} we describe the HVS Survey dataset and the MW-LMC model, as well as the orbit integration methods. In Section \ref{sec:inverse} we rewind the orbits of HVS Survey stars to determine their likely origin. From these orbits, we conduct a hypothesis test for each star to determine its origin in the Galactic Center versus the LMC. In Section \ref{sec:forward} we construct a forward model that incorporates the Hills mechanism and the selection function of the HVS Survey. We use this forward model to generate mock realizations of an HVS Survey dominated by Galactic disk runaways, LMC disk runaways, and a SMBH in the LMC. We show that the Leo Overdensity is only replicated by the SMBH of the LMC. In Section \ref{sec:lmc*}, we constrain the mass of the LMC SMBH by modeling the ejection velocities and count ratios of HVSs originating from the LMC and the Galaxy. We finally conclude and discuss our results in Section \ref{sec:discussion}.

\section{Methods}\label{sec:methods}

We analyze the complete HVS Survey data \citep{Brown2014} which includes 21 unbound late B-type main sequence stars with well-measured atmospheric properties including effective temperature, surface gravity, and projected rotation velocity $v\sin i$. Almost all the stars are fast rotators, which makes it highly unlikely that these stars are horizontal branch stars \citep[e.g.][]{Kreuzer2020}. On these grounds, \citet{Brown2014} use main sequence stellar evolution tracks \citep{marigo08, girardi04} to estimate HVS masses and luminosities. The stars span Galactocentric radii of $40-120\text{ kpc}$ and stellar masses of $2.5-4.2M_{\odot}$. We supplement these 21 stars with \textit{Gaia} DR3 proper motions \citep{GaiaCollaboration2021}, and compile the kinematic data in Table \ref{tab:hvs_data}.

As a fiducial model for the Milky Way (MW)–Large Magellanic Cloud (LMC) system, we adopt the simulation presented in \citet[][hereafter GC19]{nico19}, which has proven successful in recent applications such as predicting global stellar density wakes in the halo \citep{conroy21} and a radial velocity dipole in the outer halo \citep[e.g.,][]{bystrom24, chandra24}. We specifically use ``Simulation 7'', which has an LMC of mass $M_{200}=1.47 \times10^{11} M_{\odot}$ and Milky Way of mass $M_{200}=1.03 \times10^{12} M_{\odot}$. The LMC is just past its first pericentric passage in this model \citep{besla07}. Further details can be found in Section 3 of \citetalias{nico19}; Simulation 7 is one of the two models found to best match observations.

To enable rapid calculations of orbits that allow us to vary model parameters, we extract the center-of-mass trajectories of the MW and LMC from this simulation. We then approximate the time-dependent gravitational potential of the MW-LMC system by positioning their equilibrium potentials at the respective center-of-mass locations at each time step. This approximation does not account for the perturbations in the shape of the potential; however, since HVSs are extremely fast-moving and short-lived, their orbits are relatively insensitive to such perturbations. For example, \cite{boubert20} explored the impact of perturbations to the shape of the Galactic potential due to the LMC, and found that any resulting deflections of HVS trajectories are smaller than the uncertainties in \textit{Gaia} proper motions. In addition, the LMC orbit in the last $400\text{ Myr}$ is only weakly affected by the mass of the LMC, as discussed by \citetalias{nico19}. The equilibrium potentials of the LMC and MW, along with the orbit integrations in the time-varying potential, are implemented in \texttt{gala} \citep{galajoss} with a fixed timestep of $0.1\,\text{Myr}$ using a Leapfrog integrator. Over this interval, a star moving at $1000\,\rm km\,s^{-1}$ travels $0.1\,\text{kpc}$, which is negligible compared to the typical spatial position uncertainties of $\sim10\,\text{kpc}$. Our results are not sensitive to variations in the integration timestep. Both our inverse and forward models have a total integration time of $400\text{ Myr}$. The Galactic potential is based on the \texttt{MilkyWayPotential} class \citep{galajoss, bovy15} which includes a disk, bulge, nucleus, and halo. We modify the halo mass to make the total mass of the Milky Way  $10^{12}M_{\odot}$. The LMC potential has a total mass of $1.5\times10^{11}M_{\odot}$ to match the \citetalias{nico19} simulation. We assume an LMC stellar disk mass of $2.5\times10^9 M_{\odot}$ \citep{kim98} with a scale radius of $1.5\text{ kpc}$ and scale height of $0.5 \text{ kpc}$, and we orient the disk according to the position and inclination angles measured by \cite{vandermarel14}. We assume that the disks of the Galaxy and the LMC do not change their orientation significantly in the last 400 Myr.

\begin{table*}
\centering
\caption{HVS Survey Data with \textit{Gaia} DR3 astrometry. Velocities are in $\text{km}\text{ s}^{-1}$ and proper motions are in $\text{mas}\text{ yr}^{-1}$. All of these stars have RUWE values close to 1. We describe how $p-$values and the likelihood ratios are calculated for each origin hypothesis in Section \ref{sec:inverse}.}
\label{tab:hvs_data}
\begin{tabular}{rrlllllllll}
\toprule
 HVS &   source\_id &  $v_\text{helio}$ &             pmra &            pmdec &  $d_\text{helio}$ [kpc] & $p_{\text{MW}}$ & $p_{\text{LMC}}$ & $ \log \frac{\mathcal{L_{\text{LMC}}}}{\mathcal{L}_{\text{MW}}} $ \\
\midrule
   1 &  577294697514301440 &  831.10 $\pm$ 5.70 & -0.60 $\pm$ 0.60 & -0.47 $\pm$ 0.39 & 102.24 $\pm$ 14.60 & 0.72 & 0.12 & -0.49 \\
   4 &  699811079173836928 &  600.90 $\pm$ 6.20 & -0.20 $\pm$ 0.26 & -0.60 $\pm$ 0.19 &   63.80 $\pm$ 9.70 & 0.92 & 0.01 & -2.6 \\
   5 & 1069326945513133952 &  545.50 $\pm$ 4.30 &  0.00 $\pm$ 0.08 & -0.99 $\pm$ 0.11 &   44.20 $\pm$ 5.09 & 0.00 & 0.00 & nan \\
   6 & 3867267443277880320 &  609.40 $\pm$ 6.80 &  0.12 $\pm$ 0.30 &  0.12 $\pm$ 0.23 &   55.36 $\pm$ 6.88 & 0.00 & 0.46 & 7.0 \\
   7 & 3799146650623432704 &  526.90 $\pm$ 3.00 & -0.09 $\pm$ 0.18 &  0.02 $\pm$ 0.13 &   52.17 $\pm$ 6.25 & 0.00 & 0.49 & 27.0 \\
   8 &  633599760258827776 &  499.30 $\pm$ 2.90 & -0.88 $\pm$ 0.16 & -0.28 $\pm$ 0.14 &   53.19 $\pm$ 9.80 & 0.00 & 0.00 & -53.0 \\
   9 & 3830584196322129920 &  616.80 $\pm$ 5.10 &  0.26 $\pm$ 0.43 & -0.81 $\pm$ 0.65 &  74.10 $\pm$ 11.60 & 0.44 & 0.77 & -0.13 \\
  10 & 3926757653770374272 &  467.90 $\pm$ 5.60 & -1.09 $\pm$ 0.45 & -0.99 $\pm$ 0.21 &   51.76 $\pm$ 5.72 & 0.57 & 0.80 & -0.38 \\
  12 & 3809777626689513216 &  552.20 $\pm$ 6.60 &  0.93 $\pm$ 0.88 & -0.19 $\pm$ 0.58 &   64.83 $\pm$ 8.36 & 0.17 & 0.33 & 0.18 \\
  13 & 3804790100211231104 &  569.30 $\pm$ 6.10 &  0.07 $\pm$ 0.79 & -0.20 $\pm$ 0.63 & 105.58 $\pm$ 19.45 & 0.54 & 0.42 & 0.098 \\
  14 & 3859275333773935488 &  537.30 $\pm$ 7.20 & -2.17 $\pm$ 1.38 &  2.28 $\pm$ 1.68 & 102.66 $\pm$ 16.55 & 0.00 & 0.00 & 0.55 \\
  15 & 3794074603484360704 &  461.00 $\pm$ 6.30 & -1.30 $\pm$ 0.36 & -0.48 $\pm$ 0.23 &   66.16 $\pm$ 9.75 & 0.02 & 0.45 & 0.15 \\
  16 & 3708104343359742848 &  429.80 $\pm$ 7.00 & -1.29 $\pm$ 0.50 & -0.54 $\pm$ 0.29 &  70.93 $\pm$ 11.43 & 0.37 & 0.73 & -0.22 \\
  17 & 1407293627068696192 &  250.20 $\pm$ 2.90 & -1.13 $\pm$ 0.09 & -0.93 $\pm$ 0.10 &   49.82 $\pm$ 3.90 & 0.00 & 0.00 & nan \\
  18 & 2872564390598678016 &  237.30 $\pm$ 6.40 &  0.01 $\pm$ 0.34 & -0.24 $\pm$ 0.32 &  77.34 $\pm$ 10.68 & 0.71 & 0.00 & -1.0 \\
  19 & 3911105521632982400 & 592.80 $\pm$ 11.80 &  0.52 $\pm$ 1.08 & -0.98 $\pm$ 1.14 &  97.32 $\pm$ 15.24 & 0.23 & 0.14 & 0.29 \\
  20 & 3800802102817768832 &  512.10 $\pm$ 8.50 & -0.18 $\pm$ 0.66 & -0.99 $\pm$ 0.56 &  75.40 $\pm$ 10.76 & 0.60 & 0.71 & 0.14 \\
  21 &  834069905715968640 &  356.80 $\pm$ 7.50 & -0.20 $\pm$ 0.41 & -0.65 $\pm$ 0.65 &  72.11 $\pm$ 13.95 & 0.85 & 0.66 & -0.59 \\
  22 & 3897063727354575488 & 597.80 $\pm$ 13.40 &  0.07 $\pm$ 0.87 & -0.59 $\pm$ 0.67 &  83.98 $\pm$ 13.15 & 0.62 & 0.65 & 0.17 \\
  23 & 2681450921590663296 &  259.30 $\pm$ 9.80 & -1.21 $\pm$ 1.29 & -2.46 $\pm$ 1.50 & 114.87 $\pm$ 20.10 & 0.00 & 0.00 & -0.26 \\
  24 & 3810351984075984768 &  492.50 $\pm$ 5.30 &  0.10 $\pm$ 0.31 & -0.40 $\pm$ 0.26 &   54.08 $\pm$ 7.47 & 0.05 & 0.69 & 0.45 \\
\bottomrule
\end{tabular}
\end{table*}

\section{Inverse Modeling The HVS Survey}\label{sec:inverse}

Here, we investigate the origin of the HVS Suvey stars by integrating their orbits back in time. For each star, we take 10,000 samples from its radial velocity, distance, and proper motion uncertainties. The distances are inferred from fitting the observed temperatures and surface gravities with stellar models \citep{Brown2014}, since {\it Gaia} parallaxes are unconstraining at the stars' large distances. Comparable distances have been calculated by \citet{Kreuzer2020}.

For each trajectory, we record the closest approach of the HVS to the Galactic Center and the LMC Center. The resulting distribution of closest passages is approximated as a 3D Gaussian, from which we compute the Mahalanobis distance \citep{mahalanobis36}---a statistical measure that calculates the distance between a point (the LMC/MW galactic center) and a distribution (the distribution of closest passages). Since the Mahalanobis distance follows a \( \chi^2 \) distribution with three degrees of freedom, we can derive a \( p \)-value for the hypothesis that the HVS passes through the Galactic Center or LMC Center. Additionally, we compute a likelihood ratio to compare the two origin hypotheses in cases where both are consistent with the data.

\begin{figure*}
    \centering
    \includegraphics[width=0.8\linewidth]{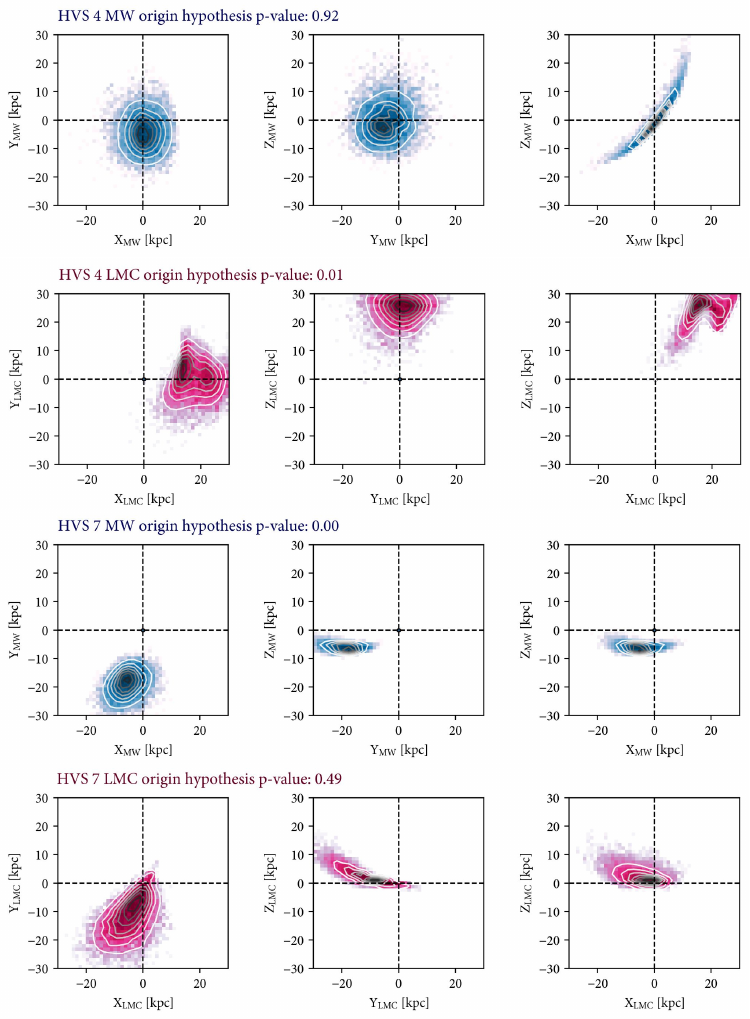}
    \caption{Hypothesis tests for origin scenarios at the Galactic Center (GC) and the Large Magellanic Cloud (LMC). For each hypervelocity star (HVS), we sample the observational uncertainties in its present-day 6D phase-space location and compute its closest passage to the GC (blue) and LMC center (red). The resulting distributions of closest passages are shown as histograms, with KDE approximations overplotted as grey/white contours. We evaluate the likelihood of each origin scenario by calculating the Mahalanobis distance to the respective centers and derive a p-value. The results for HVS4 and HVS7 are presented: while HVS4 is consistent with an origin at the Galactic Center, HVS7 is not. Instead, HVS7’s trajectory is only consistent with an origin from the LMC center.}
    \label{fig:pvals}
\end{figure*}

Figure \ref{fig:pvals} shows the result of this analysis for two stars, HVS 4 and HVS 7. The blue histograms show the nearest approach to the Galactic Center, and the magenta histograms show the nearest approach to the LMC center. We also overplot Gaussian kernel density estimates (KDEs) of each histogram as grey/white contours. HVS 4 is consistent with originating from the Galactic Center, and is inconsistent with coming from the LMC center. On the other hand, HVS 7 is inconsistent with a Galactic Center origin---it is nearly $30\text{ kpc}$ offset from passing through the Galactic Center, which is also several scale radii away from the Galactic disk. Instead, it is consistent with an origin from the LMC center.

\begin{figure}
    \centering
    \includegraphics[width=\linewidth]{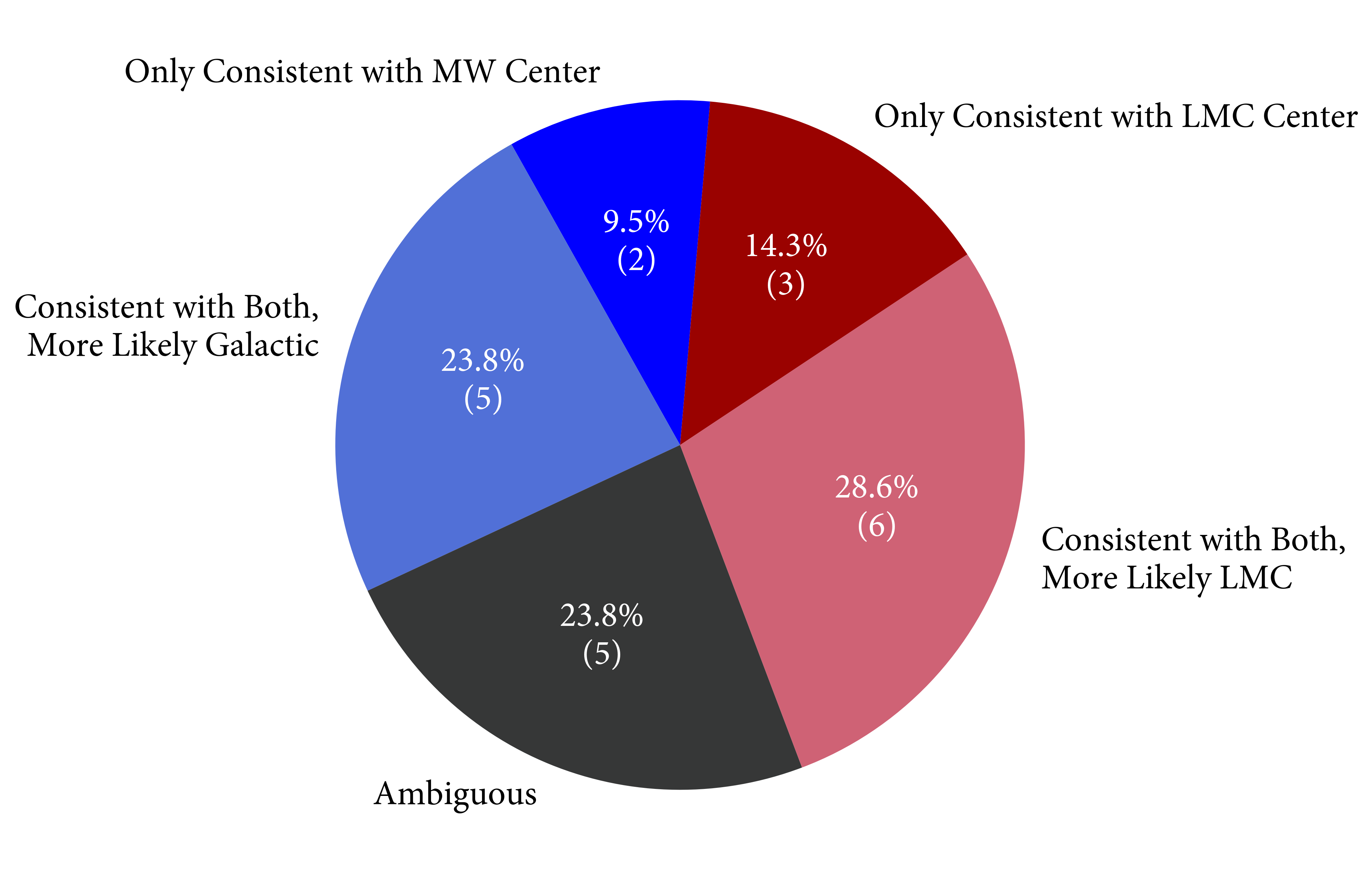}
    \caption{Classification of HVSs based on their likely origins. The pie chart categorizes the stars into five groups: those only consistent with an origin at the Galactic Center (dark blue), those only consistent with the LMC center (dark red), stars consistent with both origins but more likely from Sgr A* (light blue), stars consistent with both origins but more likely from LMC* (light red), and ambiguous cases where neither origin can be definitively favored (grey). Among the HVSs that can be confidently classified, 9 out of 16 stars originate from the LMC center. The $p$-values and likelihood ratios are summarized in Table \ref{tab:hvs_data}.}
    \label{fig:pie}
\end{figure}

We perform this analysis for all 21 HVS Survey stars, and summarize the result in Table \ref{tab:hvs_data} and Figure \ref{fig:pie}. In the pie chart, dark blue indicates that the star is consistent with a Galactic Center origin but not an LMC center origin $(p_{\text{MW}}>0.05) \wedge (p_{\text{LMC}}<0.05)$, while dark red indicates that the star is only consistent with an LMC center origin $(p_{\text{MW}}<0.05) \wedge (p_{\text{LMC}}>0.05)$. Lighter blue/red indicates stars that are consistent with both $(p_{\text{MW}}>0.05) \wedge (p_{\text{LMC}}>0.05)$, but are more likely to be from the Galactic Center (light blue) or the LMC center (light red) based on their likelihood ratios. In Table \ref{tab:hvs_data}, a positive $\log \mathcal{L}_{\text{LMC}}/\mathcal{L}_{\text{MW}}$ indicates that an LMC origin is preferred, while negative values indicate that a MW origin is preferred. 5 stars are consistent with neither centers $(p_{\text{MW}}<0.05) \wedge (p_{\text{LMC}}<0.05)$, and are indicated in grey. These stars are also inconsistent with a supernovae runaway origin, as their relative velocities to the disk are either beyond $800\text{ km}\text{ s}^{-1}$, they impact the disk at distances greater than $20\text{ kpc}$ where star formation should be negligible, or the difference between the lifetime of the HVS and its travel time from the disk is large enough to rule out a SN-progenitor massive companion \citep{brown12_hvs5, brown13_hvs17}. A deeper understanding of the LMC-MW orbit and improved \textit{Gaia} DR4 data should reveal clearer insights into the origin of these stars. Among the HVSs that can be confidently classified, slightly more than half (9 out of 16) are most likely to originate in the LMC center. This analysis is similar to what has been done by \cite{Brown2018}, with the exception that we add the possibility of an LMC origin. Using lower-precision proper motions from {\it Gaia} DR2, \cite{Brown2018} found that two stars (HVS 17 and HVS 7) are inconsistent with a Galactic Center origin---we find that HVS 7 is consistent with a LMC center origin, while HVS 17 remains ambiguous in its origin.

\subsection{The Leo Overdensity}

\begin{figure}
    \centering
    \includegraphics[width=\linewidth]{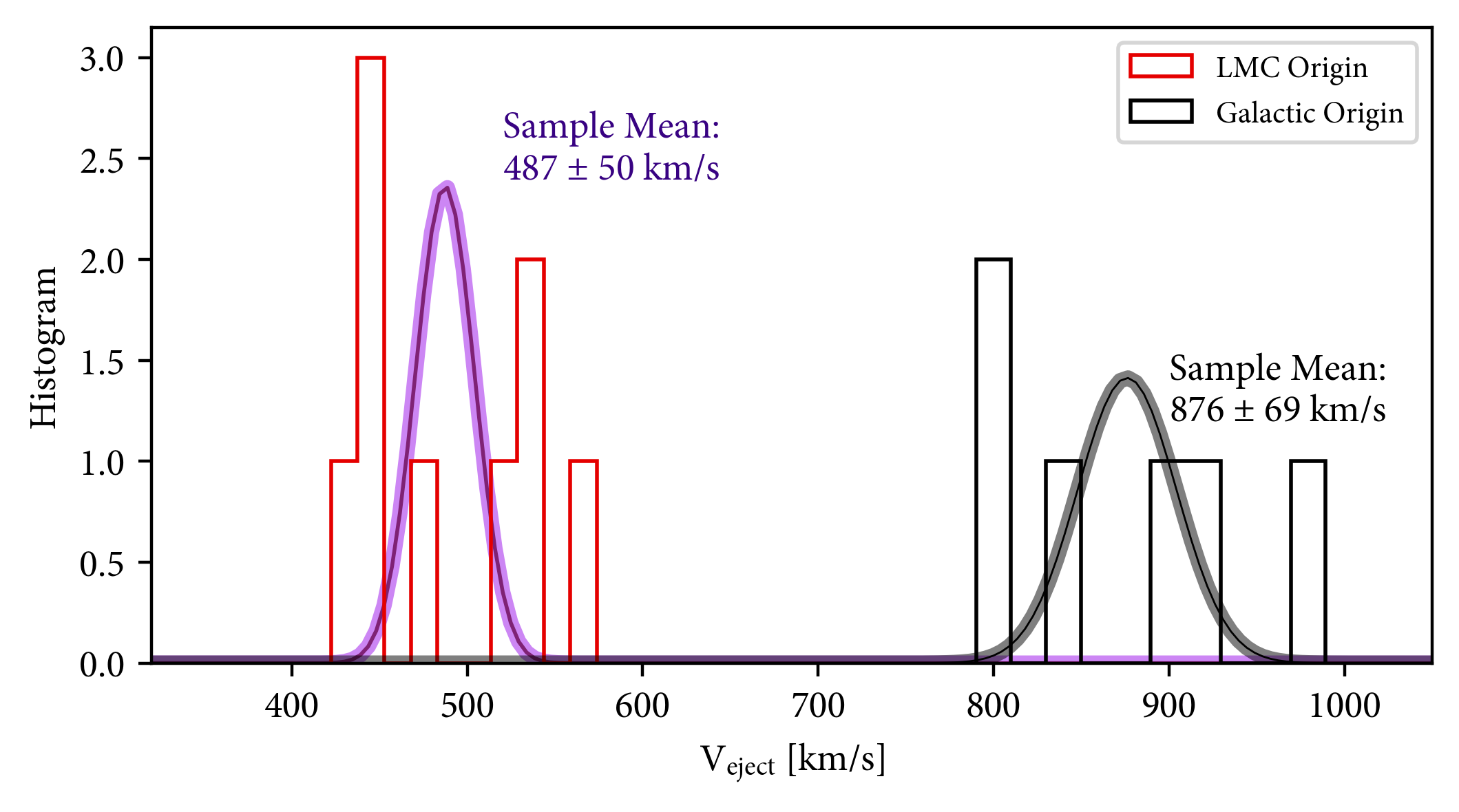}
    \caption{Mean ejection velocities of HVS from LMC* and Sgr A*. We show the individual ejection velocities as a histogram, and plot the sample mean and uncertainty on the sample mean as purple (LMC origin) and black (Galactic origin) gaussian curves.}
    \label{fig:obs_veject}
\end{figure}

In the top left panel of Figure \ref{fig:fourpanel}, we color the HVS Survey stars by their classified origin. The Leo Overdensity nearly entirely consists of LMC-origin stars, confirming the prediction of \citet{Boubert2016}. We will further explore this curious result (along with the rest of the figure) in Section \ref{sec:forward}. In Figure \ref{fig:obs_veject}, we show a histogram of ejection velocities of HVS from their respective origins. The ejection velocity is calculated as the relative velocity of the HVS to the LMC/Galaxy center-of-mass at the time of closest approach. We calculate this velocity for orbits that pass within $1\text{ kpc}$ of the respective galaxy centers, and record the mean value of those velocities as the ejection velocity. We note that the LMC in our model does not include a bulge component, and the ejection velocities at the LMC Center versus at the LMC disk scale radius only differ by $\sim30\text{ km}\text{ s}^{-1}$.

The systematically lower ejection velocities of LMC-origin HVSs suggest that the ejection mechanism is galaxy mass-dependent. The Hills mechanism is consistent with this observation, as a hypothetical SMBH in the LMC center would likely have a smaller mass than Sgr A*. In contrast, disk runaway stars would not show this correlation, as their ejection velocities are independent of galaxy mass. In Section \ref{sec:forward}, we show that if Galactic disk runaways were produced at the observed LMC ejection velocities, they would be readily detectable by the HVS Survey and would be distributed all across the sky.

Building on these findings, we explore the hypothesis that the LMC-origin HVS are produced by an SMBH at the center of the LMC via the Hills mechanism. In deference to the naming of other supermassive black holes, we refer to this black hole as LMC*. In the following section, we investigate if  LMC* could self-consistently account for the observed properties of the HVSs Survey sample.

\begin{figure*}
    \centering
    \includegraphics[width=0.7\linewidth]{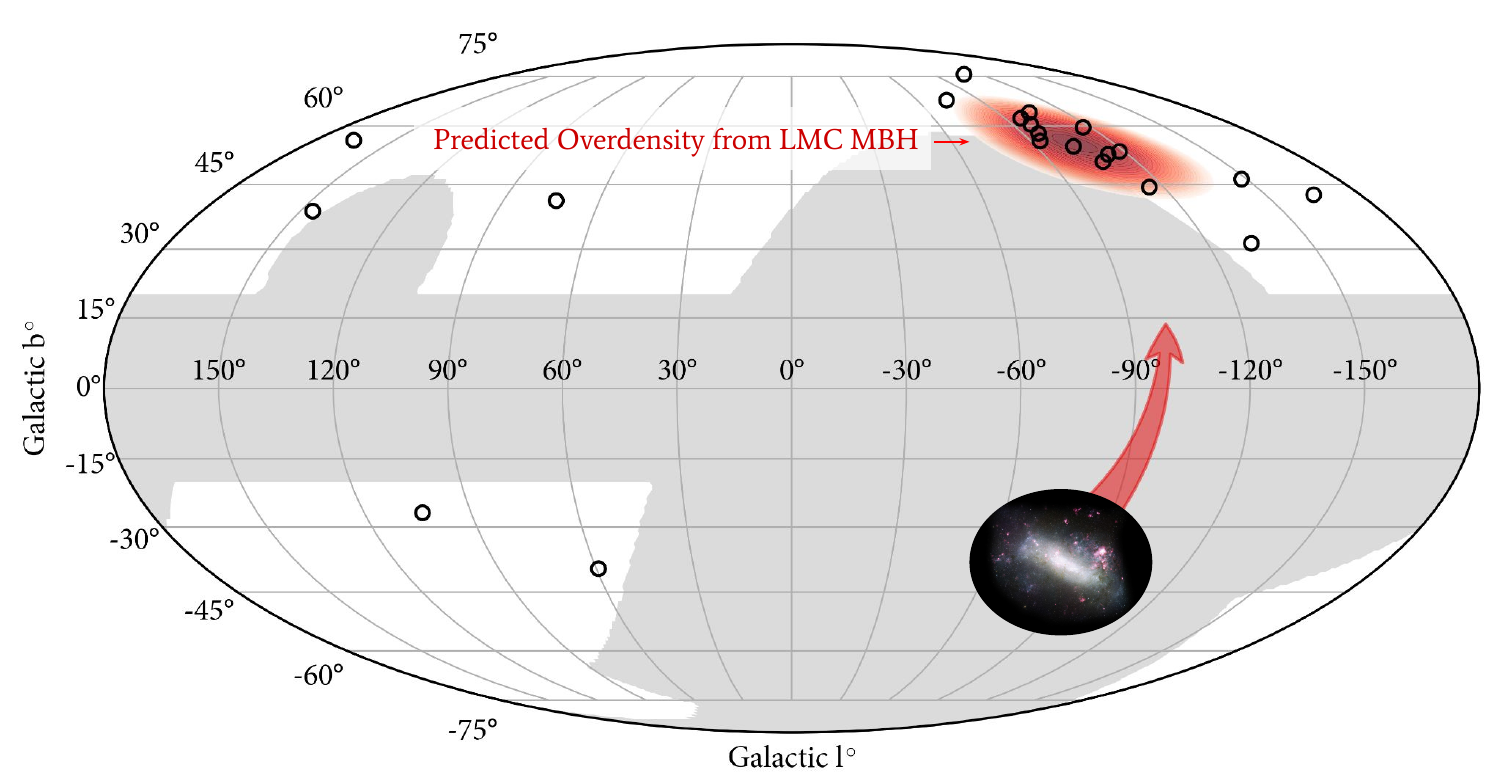}
    \caption{
    Predicted on-sky overdensity of hypervelocity stars originating from a $6\times10^5 M_{\odot}$ supermassive black hole in the LMC. The black open circles denote the Galactic coordinates of hypervelocity stars detected in the HVS Survey, while the grey-shaded regions mark areas excluded from the survey. The current position of the LMC is illustrated with a representative image, and its orbital trajectory is drawn with a red arrow. The forward model incorporating an SMBH in the LMC along with the selection effects of the HVS Survey predicts a prominent overdensity of HVSs in the region enclosed by the red contours. The overdensity arises because stars are boosted in the direction of the LMC's orbit. This model accurately reproduces the observed overdensity location, supporting the hypothesis of an SMBH in the LMC as a source of these stars.
    }
    \label{fig:onsky}
\end{figure*}

\begin{figure*}[t]
    \centering
    \includegraphics[width=\linewidth]{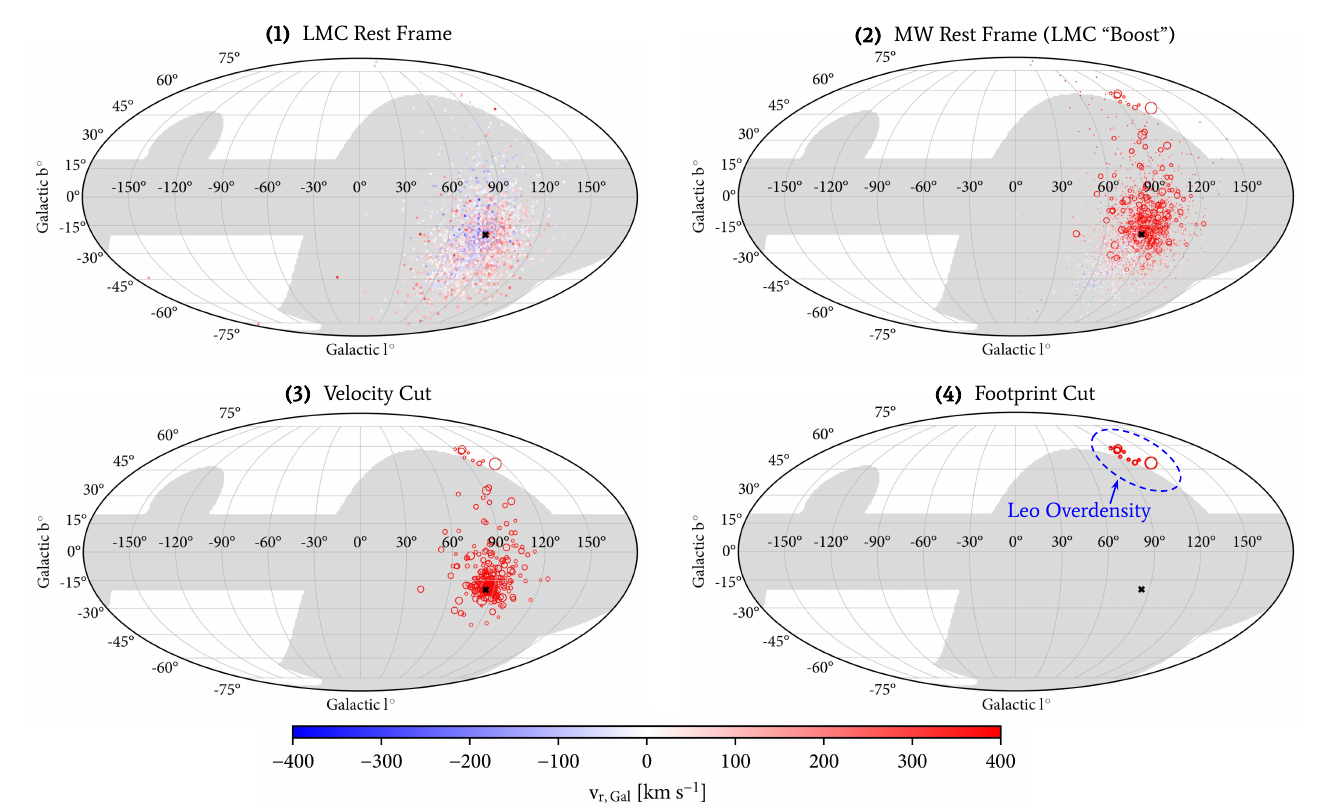}
    \caption{How hypervelocity stars make it into the HVS Survey. In the first panel, we show the LMC rest-frame velocities of stars ejected from a $6\times10^5M_{\odot}$ black hole via the Hills mechanism. We only show stars that pass the HVS Survey color-magnitude cuts, and the \textbf{$\times$} marks the present-day location of the LMC center in the model. The Galactocentric radial velocities $v_{r,\text{Gal}}$ are randomly scattered, with most stars being ejected at modest velocities. In the second panel, we show the velocities of these stars in the rest-frame of the Galaxy. The size of each point is proportional to the excess velocity over the local Galactic escape velocity. Only the stars that are ejected parallel to the orbit of the LMC are boosted beyond the local escape velocity. In the third panel, we select stars that are above the escape velocity, revealing a trail of hypervelocity stars leading ahead of the LMC orbit. In the last panel, we show stars that make it into the HVS Survey footprint. The leading tip of LMC hypervelocity stars manifests in the data as the Leo Overdensity.}
    \label{fig:selection}
\end{figure*}

\vspace{8mm}

\section{Forward Modeling The HVS Survey}\label{sec:forward}

In this Section, we forward model the HVS Survey. Our goal is to determine whether the observed HVS distribution can be explained by ejection from a SMBH in the LMC (hereafter LMC*) via the Hills mechanism.

\subsection{Hills Mechanism \& The Selection Function}
The main ingredients of the Hills mechanism are: (1) the mass of LMC*, (2) binary star masses, (3) binary separations $a_{\text{bin}}$ prior to tidal disruption, and (4) pericenter distances of the binary orbit around the SMBH, $r_{\text{peri}}$. The Hills mechanism takes these ingredients as input and outputs both the ejection velocity and the ejection probability for a given star. We note that ``ejection velocity'' here, and throughout the paper, is the asymptotic velocity of the star as it escapes the SMBH's sphere of influence. This is significantly lower than the star’s initial velocity immediately after being tidally released, which can reach extreme values on the order of \( 10,000 \, \rm km\,s^{-1} \) \citep[e.g.][]{Hills1988}.

Following \citet{Bromley2006} and \citet{Kenyon2008}, we model the production of HVSs using the equations:
\begin{align}\label{eq:hills}
v_{\text{ej}} &= 1,370\,\text{km\,s}^{-1} 
\left( \frac{a_{\text{bin}}}{0.1\,\text{AU}} \right)^{-1/2} 
\left( \frac{m_b}{M_\odot} \right)^{1/3} \nonumber \\
&\quad \times \left( \frac{M}{4 \times 10^6\,M_\odot} \right)^{1/6} f_R,
\end{align}

\begin{align}\label{eq:hills_D}
D &= \left( \frac{r_{\text{peri}}}{a_{\text{bin}}} \right) 
\left( \frac{10^6}{M} \frac{m_b}{2} \right)^{1/3},
\end{align}

\begin{multline}\label{eq:hills_fr}
f_R = 0.774 + 
\left( 0.0204 + 
\left\{ -6.23 \times 10^{-4} + 
\left[ 7.62 \times 10^{-6} \right.\right. \right. \\
\quad + \left( -4.24 \times 10^{-8} + 8.62 \times 10^{-11} D \right) D \\
\left.\left.\left. \right] D \right\} D \right) D,
\end{multline}

\begin{equation}
P_{\text{ej}} =
\begin{cases} 
1 - \frac{D}{175}, & \text{if } D < 175, \\
0, & \text{otherwise.}
\end{cases}
\label{eq:hills_pej}
\end{equation}

where $v_{\text{ej}}$ is the ejection velocity, $a_{\text{bin}}$ is the binary separation, $r_{\text{peri}}$ is the pericenter distance of the binary to the SMBH, and $m_b$ is total binary mass. $v_{\rm ej}$ represents the median ejection velocity for a circular orbit, assuming random orientations and phases \citep[e.g.][]{Sari2010}. The approximation for $f_R$ was obtained by \citet{Bromley2006}, who fit a polynomial to results of three-body simulations performed by \citet{Hills1988}.

There are two regimes that describe the dynamics of scattering towards the SMBH that eventually leads to the binary disruption: the ``full loss cone'' regime and the ``empty loss cone'' regime \citep[][]{Lightman1977}. In the full loss cone regime, the SMBH does not significantly modify the distribution of periapsis in its vicinity, as two body scattering efficiently reinstates orbits with low angular momentum and low periapse that may have been disrupted. Meanwhile, in the empty loss cone regime, binaries gradually diffuse to near-radial orbits that bring them near the SMBH, and all binary disruptions are launched from the binary tidal radius ($D\approx 80$). Thus, the periapsis distribution is focused at the binary tidal radius. \citet{Rossi2014} investigate the empty loss cone regime to find that the binary separations must be weighted toward wider separations (relative to the binary population observed in the local Galactic field) to match the spectrum of ejection velocities from Sgr A*. In this study, we choose to match the distribution of binary separations to the local Galactic field, and assume the full loss cone regime. This choice is consistent with the majority of previous studies modeling the HVS population \citep[e.g.,][]{Gualandris2007, Kenyon2008, Boubert2016, Erkal2019}. The system could indeed be in the full loss cone regime if scattering is stronger than simple estimates (if, for example, some unknown massive bodies are around). However, since there is an uncertainty on which regime Sgr A* and LMC* belong to, we treat with caution the interpretation of our findings for the periapsis distribution of binaries. We detail below our modeling choices in the full loss cone regime.

\begin{figure*}[t]
    \centering
    \includegraphics[width=\linewidth]{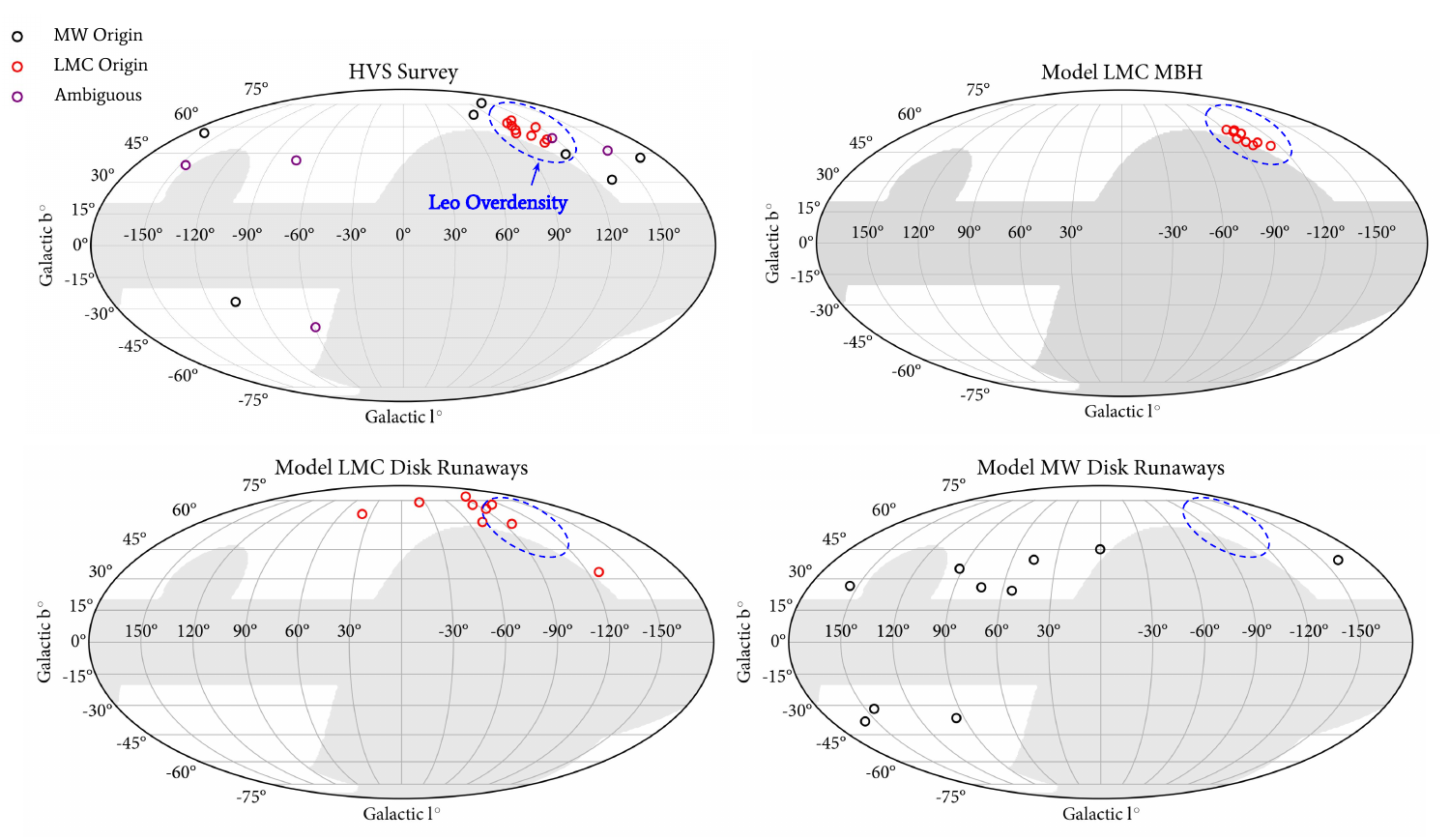}
    \caption{Mock realizations of the HVS Survey dominated by different origins: the LMC SMBH (top right), the LMC disk runaways (bottom left), or the Galactic disk runaways (bottom right). The top left panel displays the actual HVS Survey stars, colored by their origin as classified using the methods described in Section \ref{sec:inverse}. The Leo Overdensity is highlighted with dotted blue lines. For the LMC and Galactic disks, we assume constant rotation velocities of $70\text{ km}\text{ s}^{-1}$ and $200\text{ km}\text{ s}^{-1}$, respectively. The disk rotation causes the stars to be spread out across the sky, which is inconsistent with the pronounced Leo Overdensity observed in the data. Only the LMC SMBH model successfully reproduces the Leo Overdensity.
    }
    \label{fig:fourpanel}
\end{figure*}

The stellar initial mass function (IMF) produces a large number of low-mass stars; however, main-sequence stars with low masses are not observable at the distances probed by the HVS Survey and also fall outside the survey's photometric selection in color-color space.  The survey's mass cutoff is approximately $2.5\,M_\odot$. We thus restrict the primary star’s mass to $>2\,M_\odot$. In contrast, we allow the binary companion's mass to range from $0.1\,M_\odot$ to $10\,M_\odot$, following a Salpeter IMF \citep{salpeter}. While a top-heavy IMF would produce a larger fraction of higher $v_{\text{eject}}$ stars compared to a bottom-heavy IMF, only the high-$v_{\text{eject}}$ tail will be classified as a HVS by the HVS survey. As a result, the observed $v_{\text{eject}}$ distribution is largely insensitive to variations in the IMF. For this reason, we do not explore variations in the IMF in this study.

The binary separation, $a_{\text{bin}}$, balances two competing effects: smaller $a_{\text{bin}}$ values increase the ejection velocity, but simultaneously reduce the ejection probability, as shown in Equation \ref{eq:hills_pej}. This trade-off establishes practical upper and lower bounds for $a_{\text{bin}}$ relevant to the HVS Survey. If $a_{\text{bin}}$ exceeds $\sim3 \text{ AU}$, the resulting ejection velocities are too low to produce HVSs. Conversely, if $a_{\text{bin}}$ is smaller than a few hundredths of an AU, the ejection probabilities become negligible. Additionally, a physical lower limit to $a_{\text{bin}}$ exists because binary stars cannot fill each other's Roche lobe. For a $3\,M_\odot$ star, this limit corresponds to approximately $0.035\,\text{AU}$. Based on these constraints, we limit $a_{\text{bin}}$ to the range $0.03\,\text{AU}$ to $3\,\text{AU}$. Within this range, we assume that $a_{\text{bin}}$ follows a power-law distribution with an index $k_a$. We allow $k_a$ to be a free parameter between -2 and 2, with a Gaussian prior at $k_a = -1$ and standard deviation of 0.1 (i.e., a log-uniform separation distribution following Opik's law).

The pericenter distance to the SMBH, $r_{\text{peri}}$, primarily influences the probability of ejection, while also affecting the ejection velocity to a lesser extent through Equation \ref{eq:hills_fr}. Smaller $r_{\text{peri}}$ values increase the probability of binary disruption that produces HVSs. The lower bound of $r_{\text{peri}}$ is set by the stellar tidal disruption radius $r_t=r_{\text{Schwarzschild}}\times(M_{\text{BH}}/M_{\text{star}})^{1/3}$, within which the individual stars would tidally disrupt before producing a HVS. We select the higher value of $r_t$ between the binary stars to be the lower bound for $r_{\text{peri}}$. Meanwhile, the upper bound of $r_{\text{peri}}$ is set by the binary tidal disruption radius $r_{b,t}=a_{\text{bin}}\times(3M_{\text{BH}}/M_{\text{bin}})^{1/3}$, beyond which HVSs are not produced. Within this range, we parametrize the distribution of $r_{\text{peri}}$ as a power-law with an index $k_r$, where $k_r = 1$ corresponds to a constant surface density of binaries around the SMBH. While previous studies have fixed $k_r = 1$, we allow it to vary as a free parameter.

Finally, we allow for the binary system to have a delay time, $t_{\text{delay}}$, before reaching its pericentric passage around the SMBH. We implement  $t_{\text{delay}}$ by calculating the minimum age of the binary system (i.e., its age at present day if it were formed at the same time as the HVS ejection), and add a random age sampled uniformly between $0$ and $t_{\text{delay}}$. The primary effect of having a long $t_{\text{delay}}$ is that the number of surviving HVSs is reduced, since stars that have terminated their main-sequence evolution will not be detectable by the HVS survey.

At each timestep of the MW-LMC orbital evolution, we sample binaries at a prescribed rate from the IMF and the power-law distribution of binary separations. We then pass the binary properties through Equation~\ref{eq:hills}, and launch stars from the respective galaxies' rest-frame according to their ejection probabilities. The orbits of these stars are then integrated forward in time for $400\text{ Myr}$, accounting for the time-varying potential. Simultaneously, we follow their stellar evolution using a grid of MIST isochrones \citep{dotter16, choi16} to determine their stellar atmospheric properties. We finally ``observe'' the resulting population of stars from the Galactic rest frame at present day, and apply a selection function to match the observational constraints of the HVS Survey. The specific selection criteria applied are:

\begin{itemize}
    \item Present-day Galactocentric radial velocity must exceed the local escape velocity, as defined by \cite{Brown2014}:\\\( v_{\text{Gal}} > v_{\text{esc, Brown14}} \)
    \item Stars should not be evolved beyond the terminal-age main sequence (TAMS), and must have surface gravities of main-sequence stars:\\\( (\text{EEP} < \text{EEP}_{\text{TAMS}} ) \, \wedge (\log g> 4.0 ) \)\\
    \item Photometric magnitude and color cuts to match the HVS Survey.\\\( (g > 17) \, \wedge \, (g < 20.25) \, \wedge \\ \left( -0.4 < (g-r) < -0.43(u-g) + 0.18 \right)\wedge  \\ \left( -0.5 < (r-i) < 0 \right) \, \wedge \\ \left( 2.2(g-r) + 0.1 < (u-g) < 1.07 \right) \)
    \item Spatial footprint of the HVS Survey, which is inherited from the SDSS-III DR8 footprint \citep{sloan8}.
\end{itemize}

Based on the procedure described above, we show the result of forward-modeling a HVS Survey produced by an LMC* with a mass of \(6 \times 10^5 M_\odot\), \(k_a = -1\), and \(k_r = 2\) in Figure \ref{fig:onsky}. The on-sky distribution of HVSs produced by the LMC is shown as a red KDE contour, and the present-day location and orbital direction of the LMC is illustrated as a red arrow. We see that: (1) the LMC* HVSs are highly clustered, and (2) this cluster lies directly on top of the Leo Overdensity. The clustering can be explained by the $\sim 300\,{\rm km\,s^{-1}}$ velocity ``boost'' that the HVSs gain due to the LMC's center of mass motion. Stars that are ejected along the orbital direction receive a positive velocity boost, while those ejected anti-aligned with the orbit are decelerated---effectively ``frozen'' in the Galactic reference frame. We illustrate this process in Figure \ref{fig:selection}. In the first panel, we show the Galactocentric radial velocities $v_{r,\text{Gal}}$ of the stars ejected via the Hills mechanism in the LMC rest frame. We only show stars that pass the HVS Survey color-magnitude cuts (i.e., B-type main sequence stars), and the final location of the LMC in the model is marked with $\times$. In the LMC rest frame, $v_{r,\text{Gal}}$ are randomly scattered, and most stars are ejected at modest velocities. In the second panel, we shift to the Galactic rest frame. The size of each point is proportional to its excess over the local Galactic escape velocity. If the velocity is lower than the escape velocity, the points are set to the minimum size. As expected, only the stars that are parallel to the direction of the LMC orbit are boosted beyond the escape velocity. In the third panel, we apply the velocity selection function, which reveals a clear trail of HVSs in the leading direction of the LMC orbit. Finally, we apply the HVS Survey footprint in the fourth panel. The survey footprint only allows for the Northernmost tip of the HVS trail to be seen, which manifests as the Leo Overdensity in the data. In conclusion, we are able to quantitatively reproduce both the location and the angular extent of the Leo Overdensity by combining the dynamical effects of the LMC---as already investigated by \citet{Boubert2016}---and the full selection effects of the HVS Survey. If the Leo Overdensity is indeed due to the LMC HVS population, a generic prediction is that there should be a trail of hypervelocity stars in the Sourthern hemisphere, tracing all the way back to the LMC \citep[see panel 3 in Fig. \ref{fig:selection} and][]{Boubert2016,Boubert2017}.

\subsection{Could the HVS stars be disk runaways?}

Using the same forward modeling framework, we test the hypothesis that the LMC-origin HVSs are supernova runaways --  from either the Milky Way or the LMC. For these stars, we sample the ejection velocities from the observed LMC ejection velocities in Figure \ref{fig:obs_veject}, $487\pm50\text{ km}\text{ s}^{-1}$, and sample their launching sites from an exponential radial profile that matches our assumptions for the gravitational potential. We assume that the disk rotation velocities are constant at $200\text{ km}\text{ s}^{-1}$ for the Galaxy, and $70\text{ km}\text{ s}^{-1}$ for the LMC \citep{alves00}. The true LMC disk rotation can be as high as $\sim90\text{ km}\text{ s}^{-1}$ \citep{vandermarel14}. We assume that the disks' orientations do not change over the duration of our simulation. If the disks do precess significantly over $400 \text{ Myr}$, it would impart an extra scatter in the launch velocities of runaway stars. Once these stars are launched, we follow their evolution the same way as the SMBH simulation.

Summarizing these results, Figure \ref{fig:fourpanel} presents mock realizations of the HVS Survey that is dominated by LMC* (top right), LMC disk runways (bottom left), and Galactic disk runaways (bottom right). We have assumed the same production rate ($2\text{ Myr}^{-1}$) of HVSs in each of these simulations. In the top left panel, we show the actual HVS Survey colored by the respective origins of stars. We mark the location of the Leo Overdensity with a blue dashed ellipse.

Figure \ref{fig:fourpanel} demonstrates two key points. First, if Galactic disk runaways are ejected at $400-600\text{ km}\text{ s}^{-1}$, they should be readily detectable in the HVS Survey as unbound stars, scattered all across the sky. The disk runaway stars are boosted by the disk rotation, which allows them pass the HVS Survey selection criteria even though they are ejected at lower velocities than HVSs from Sgr A*. In addition, these runaway stars do not feel the deceleration from the Galactic bulge, which can be greater than $100\text{ km}\text{ s}^{-1}$ for stars initially traveling at $1000\text{ km}\text{ s}^{-1}$ through the inner $1\text{ kpc}$. Therefore, the paucity of Galactic disk runaways ejected at $400-600\text{ km}\text{ s}^{-1}$ in the unbound HVS Survey sample suggests that their production rate is far lower than that of the Hills mechanism (this is consistent with findings from previous studies such as \citealt{Boubert2017}).

The {\it predicted} birth rate of $\sim 500\,{\rm km\,s^{-1}}$ supernova runaways is quite uncertain, depending on several poorly calibrated parameters in binary evolution \citep[e.g.][]{Evans2020}. However, the small number of observed HVSs outside the Leo Overdensity places a tight empirical limit on their birth rate. If $\sim 500\,{\rm km\,s^{-1}}$ 
 supernova runaways were launched at a sufficiently high rate to explain the Leo Overdensity with LMC disk runaways, then a larger population of MW disk runaways would swamp the HVS Survey sample. The production rate of disk runaways is proportional to the star formation rate. Since the MW forms stars at approximately ten times the rate of the LMC ($1.9M_{\odot}\text{ yr}^{-1}$ \citealt{chomiuk11} versus $0.2M_{\odot}\text{ yr}^{-1}$, \citealt{harris09}), the contribution of disk runaways from the LMC to the HVS population should be even less significant than that from the Galaxy. We thus conclude that LMC-origin HVSs cannot be primarily produced from disk runway mechanisms. This argument applies to both \citet{Blaauw1961} kicks from supernovae in binaries and dynamical launching scenarios from clusters or triples. 

Moreover, the on-sky distribution of HVSs bears strong imprints of their launching mechanism. In the case of HVSs produced by LMC*, we are able to reproduce the Leo Overdensity in the observed direction, as already shown in Figure \ref{fig:onsky}. Meanwhile, LMC disk runaways are more scattered on the sky compared to the data. This is due to the rotation of the LMC disk, which imparts a spread in tangential velocity that is twice the rotation speed of the disk. In a typical LMC-origin HVS travel time of $200\text{ Myr}$, a tangential velocity difference of $2\times70\text{ km}\text{ s}^{-1}$ imparts a spread of $\sim33^{\circ}$. Thus, the narrow extent of the Leo Overdensity ($20^{\circ}$ in the longest direction) excludes HVS production mechanisms that are spread out by disk rotation. In the case of Galactic disk runaways, we do not see a significant overdensity in any direction. In summary, Figures \ref{fig:selection}-\ref{fig:fourpanel} demonstrate that the Leo Overdensity arises from the interplay between LMC* and the HVS Survey footprint, leading to the prediction that a Southern HVS Survey with a telescope such as the Rubin Observatory should detect an excess of HVSs along the LMC's orbital trajectory.

Yet another possibility is that the LMC-origin HVSs are directly formed out of the leading arm of the HI Magellanic stream as it interacts with the Galactic disk, as evidenced by the existence of the young stellar association \textit{Price-Whelan 1} \citep{price-whelan19, nidever19}. However, in this scenario, the star-forming region should be approximately co-moving with the Magellanic stream, while the observed HVSs have an excess velocity of $400-600\text{ km}\text{ s}^{-1}$ with respect to the stream. In addition, the distance to the HVSs span $50-70\text{ kpc}$ while the distance to \textit{Price-Whelan 1} is $\sim30\text{ kpc}$. We thus conclude that the LMC-origin HVSs are unlikely to have formed directly out of the Magellanic stream.

\subsection{Further evidence of LMC* from HE 0437-5439}

Here we present additional evidence supporting an SMBH in the LMC, independent of the above results. As discussed in Section~\ref{sec:intro}, the trajectory of the southern HVS HE 0437-5439 (which is not in our sample) traces directly back to the LMC. From its radial velocity and {\it Gaia} DR2 proper motions, \citet{Erkal2019} inferred that the star was ejected from the LMC $\sim$21 Myr ago with velocity $v_{\rm ejection}= 870_{-66}^{+69}\,{\rm km\,s^{-1}}$, which is faster than the ejection velocities we infer for any of the LMC-origin HVSs in our sample. Our own integration of the star's orbit yields very similar results.

\begin{figure}
    \centering
    \includegraphics[width=\columnwidth]{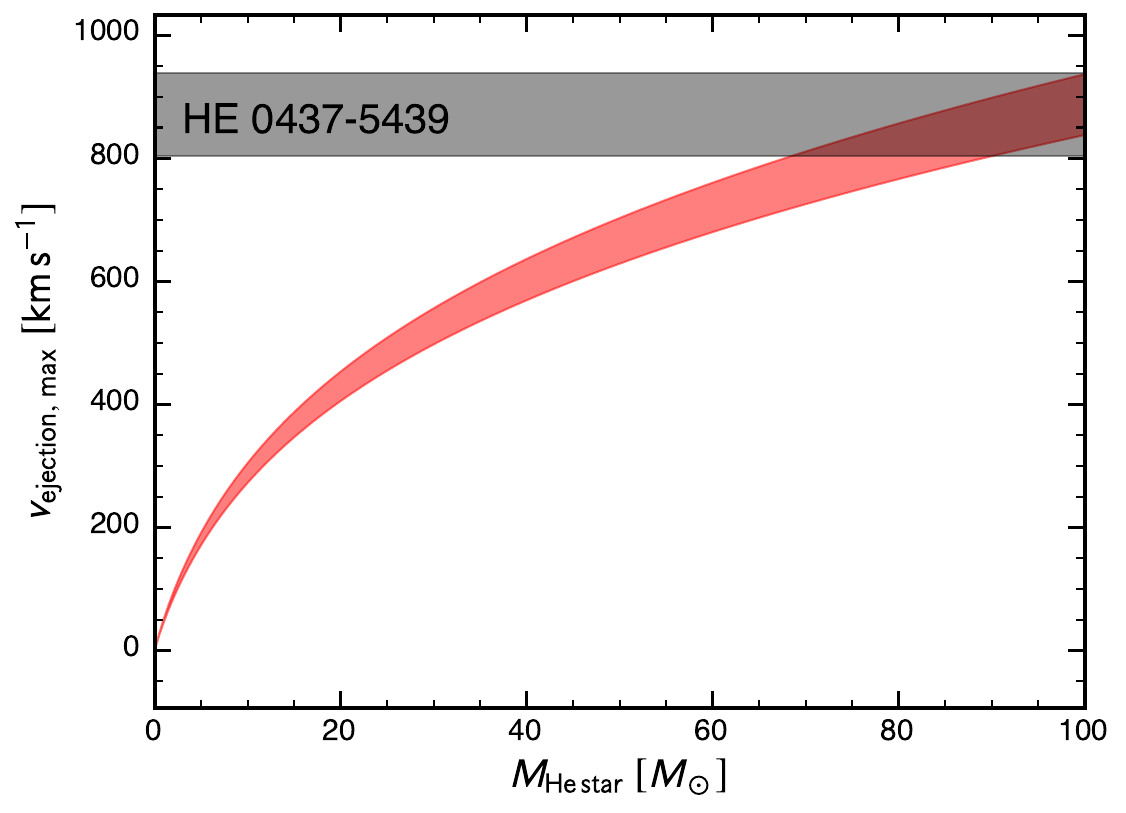}
    \caption{Gray shaded region shows constraints on the velocity at which HE 0437-5439 was ejected from the LMC \citep{Erkal2019}. Red region shows the predicted orbital velocity at Roche lobe overflow of a $9\,M_{\odot}$ with radius $4-5\,R_{\odot}$; this represents the maximum possible ejection velocity of a runaway star whose companion explodes and leaves behind no remnant. To fit inside a compact orbit, the companion would have to be a stripped helium star. Even in this case, a plausible $\lesssim 20\,M_{\odot}$ He star can only produce an ejection velocity $v_{\rm ejection} \lesssim 400\,{\rm km\,s^{-1}}$. Explaining HE 0437-5439 would require a pair-instability supernova of a $\gtrsim 80\,M_{\odot}$ He star just as its companion filled its Roche lobe. Given the implausibility of this scenario at LMC metallicity, we conclude that HE 0437-5439 was launched from a massive BH via the Hills mechanism.}
    \label{fig:hvs3_velocity}
\end{figure}

In Figure~\ref{fig:hvs3_velocity}, we consider whether this star could have been ejected from a binary via a \citet{Blaauw1961} kick (i.e., that the star is a runaway from a supernova in a binary). We consider a $9\,M_{\odot}$ main-sequence star with radius $4-5\,R_{\odot}$ in a maximally tight orbit around a denser He star, such that the $9\,M_{\odot}$ exactly fills its Roche lobe. Such a binary could form as a result of a common envelope event in which the present-day HVS ejected the envelope of a red supergiant companion and spiraled inward. The maximum ejection velocity in this case is the star's pre-SN orbital velocity \citep[e.g.][]{Bauer2021}, which is marked with a red shaded region. The Figure shows that even in this optimistic scenario, ejection velocities comparable to that inferred for HE 0437-5439 can be achieved only if the companion is a helium star with $M\gtrsim 80\,M_{\odot}$. Given that such a massive He star is unlikely to form at LMC metallicity \citep[e.g.][]{Limongi2018}, we conclude (consistent with \citealt{Erkal2019}) that HE 0437-5439 was ejected by the Hills mechanism from a massive BH. We note that the high ejection velocity of HE 0437-5439 is consistent with its exceptionally high mass. However, this star would not be included in the HVS Survey because its main-sequence lifetime is shorter than the time required to travel to the Northern Galactic hemisphere.

\begin{figure*}[ht!] 
    \centering
    \includegraphics[width=\textwidth]{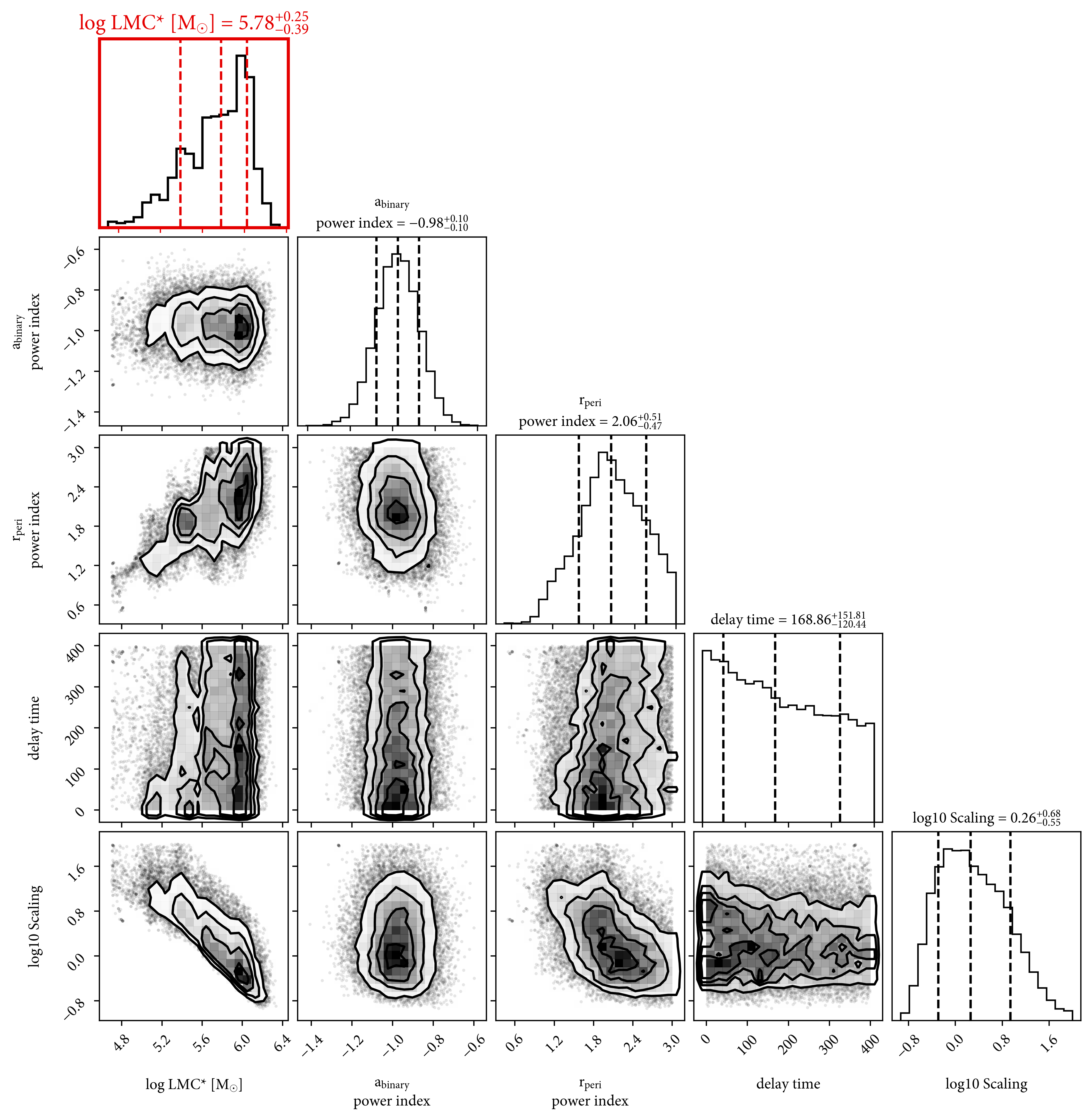} 
    \caption{MCMC sampled posterior distribution of the SMBH model parameters. The marginalized posterior distribution of each parameter is plotted as a histogram, with $M_{\text{LMC*}}$ highlighted in red. We derive an LMC* mass of $10^{5.8^{+0.2}_{-0.4}} M_{\odot}$ ($\simeq 6\times10^5 M_{\odot}$. Both $k_a$ and $t_{\text{delay}}$ are consistent with the prior, indicating that these parameters are not strongly constrained by the data. $k_r$ is constrained to be larger than 1, suggesting binaries are preferentially distributed further away from the SMBH. $\log_{10}\text{Scaling}$ peaks around 0 (i.e., same stellar densities around LMC* and Sgr A*) but has a significant spread. LMC* masses smaller than $10^5M_{\odot}$ require an unrealistically large value of $\log_{10}\text{Scaling}$.}
    \label{fig:logmass_mc}
\end{figure*}

\section{Weighing LMC*}\label{sec:lmc*}

We now estimate the mass of LMC* by assuming that the observed LMC-origin HVSs are solely produced by the Hills mechanism. We first produce a grid of simulations over a range of $M_{\text{LMC*}}$, $k_{a}$, $k_{r}$, and $t_{\text{delay}}$. $M_{\text{LMC*}}$ is sampled uniformly in log-space from $5\times10^{4}M_{\odot}$ to $3\times10^{6}M_{\odot}$ at 20 points, $k_{a}$ is sampled from $-2$ and $2$ at 10 points, $k_{r}$ from $-1$ to $3$ at 10 points, and $t_{\text{delay}}$ from $0$ to $400\text{ Myr}$ at 5 points. For each of the simulations, we measure three quantities after the selection function: the mean ejection velocity from Sgr A*, the mean ejection velocity from LMC*, and the number ratio of HVS from LMC* compared to Sgr A* (the ``count ratio''). Post-simulation, we also introduce an additional free parameter, $\log_{10}\text{Scaling}$, which allows us to arbitrarily scale up/down the production rate of LMC* HVSs compared to Sgr A* HVSs. This is equivalent to increasing/decreasing the overall stellar density around the SMBH.

Based on this grid of simulations, we conduct a parameter search as follows. We first define a likelihood function for each of the observables. For the mean ejection velocities from the respective SMBHs, we use Figure \ref{fig:obs_veject} to define a Gaussian likelihood function. For the count ratio, we adopt a Gaussian likelihood with mean $1.29$, standard deviation $0.65$, and lower truncation at $0$. This standard deviation is derived by propagating Poisson uncertainties on the LMC- and Galactic-origin HVS counts. The total likelihood function is the product of all three likelihoods. $k_a$ is given a Gaussian prior centered at $-1$ with standard deviation $0.1$, $k_r$ is given a Gaussian prior at $1$ with standard deviation $1$, $M_{\text{LMC*}}$ is given a uniform prior in log-space within the full range, $t_{\text{delay}}$ is given a uniform prior within the full range, and $\log_{10}\text{Scaling}$ is given a uniform prior from -2 to 2. We sample the posterior using \texttt{emcee} \citep{emcee} with 32 walkers and 50,000 iterations. We discard the first 200 iterations as burn-in, and thin the samples by a factor of 15.

In Figure \ref{fig:logmass_mc} we present the results from the MCMC sample. We highlight the parameter of interest, $M_{\text{LMC*}}$, in red. The marginal probability distribution of LMC* mass is $10^{5.8^{+0.2}_{-0.4}}  M_{\odot}$ ($\simeq 6\times10^5 M_{\odot}$). In the grid of simulations, we find that the mass of LMC* is the dominant parameter affecting the count ratio. While an LMC* of a wide range of masses can produce HVSs at the observed ejection velocities, the mass of LMC* determines how many HVSs the black hole can produce (given a fixed stellar density). Thus, the strong constraint on the mass of LMC* comes from the fact that nearly half of the HVSs in our sample originate from the LMC. 

Meanwhile, $k_a$ is consistent with the prior, meaning that the data are not constraining. $k_r$ peaks at 2 with a spread of 0.5, which is significantly displaced from the prior which was centered at 1. Geometrically, $k_r=1$ corresponds to a plane of orbits with uniform density, while $k_r=2$ corresponds to a sphere of orbits with uniform density. However, this interpretation hinges upon the assumption of a full loss cone (see Section \ref{sec:forward}). In the empty loss cone scenario, every binary is launched at the binary tidal radius, and the HVS data has no constraining power on the distribution of pericenters. $t_{\text{delay}}$ is not well constrained, although the data slightly prefers a smaller value. This is somewhat expected---an increased $t_{\text{delay}}$ reduces the number of surviving HVSs, and we observe numerous LMC-origin HVSs. Lastly, $\log_{10}\text{Scaling}$ slightly prefers a scaling ratio that is larger than one; $10^{0.26}\simeq1.8$. The data weakly prefers a higher density of stars near LMC* compared to Sgr A*. We note that $\log_{10}\text{Scaling}$ has a directly inverse relationship with $M_{\text{LMC*}}$: the model can accommodate a lower $M_{\text{LMC*}}$ value by scaling up the stellar density. However, to fit an $M_{\text{LMC*}}$ smaller than $10^5M_{\odot}$, the model requires the stellar density around LMC* to be unrealistically higher ($50-100\times$) than the density around Sgr A*.

\section{Conclusion \& Discussion}\label{sec:discussion}

We have investigated the origin of 21 unbound B-type main sequence stars in the HVS Survey. By using \textit{Gaia} DR3 proper motions and a state-of-the-art model for the LMC-MW motion, we integrate the orbits of these stars back in time. We construct a hypothesis test that each star originates from either the Galactic Center of the LMC center. Among the stars we can confidently classify, 7 out of 16 stars are consistent with originating from the Galactic Center, while the other 9 stars are consistent with originating from the LMC center.

The LMC-origin HVSs are clustered on the sky and show systematically lower ejection velocities that are consistent with being produced by a less massive SMBH compared to Sgr A*. Motivated by this finding, we construct a forward-model to simulate a mock HVS Survey that models the Hills mechanism and the selection function of the survey. We use this model to show that an SMBH in the LMC Center, LMC*, can self-consistently produce hypervelocity stars that match the observed distribution of locations and velocities of the HVS Survey. Specifically, this model predicts an overdensity of hypervelocity stars at precisely the location of the Leo Overdensity. These results confirm the hypothesis of \citet{Boubert2016}, which was put forth before \textit{Gaia} proper motions were available.

\begin{figure}
    \centering
    \includegraphics[width=\linewidth]{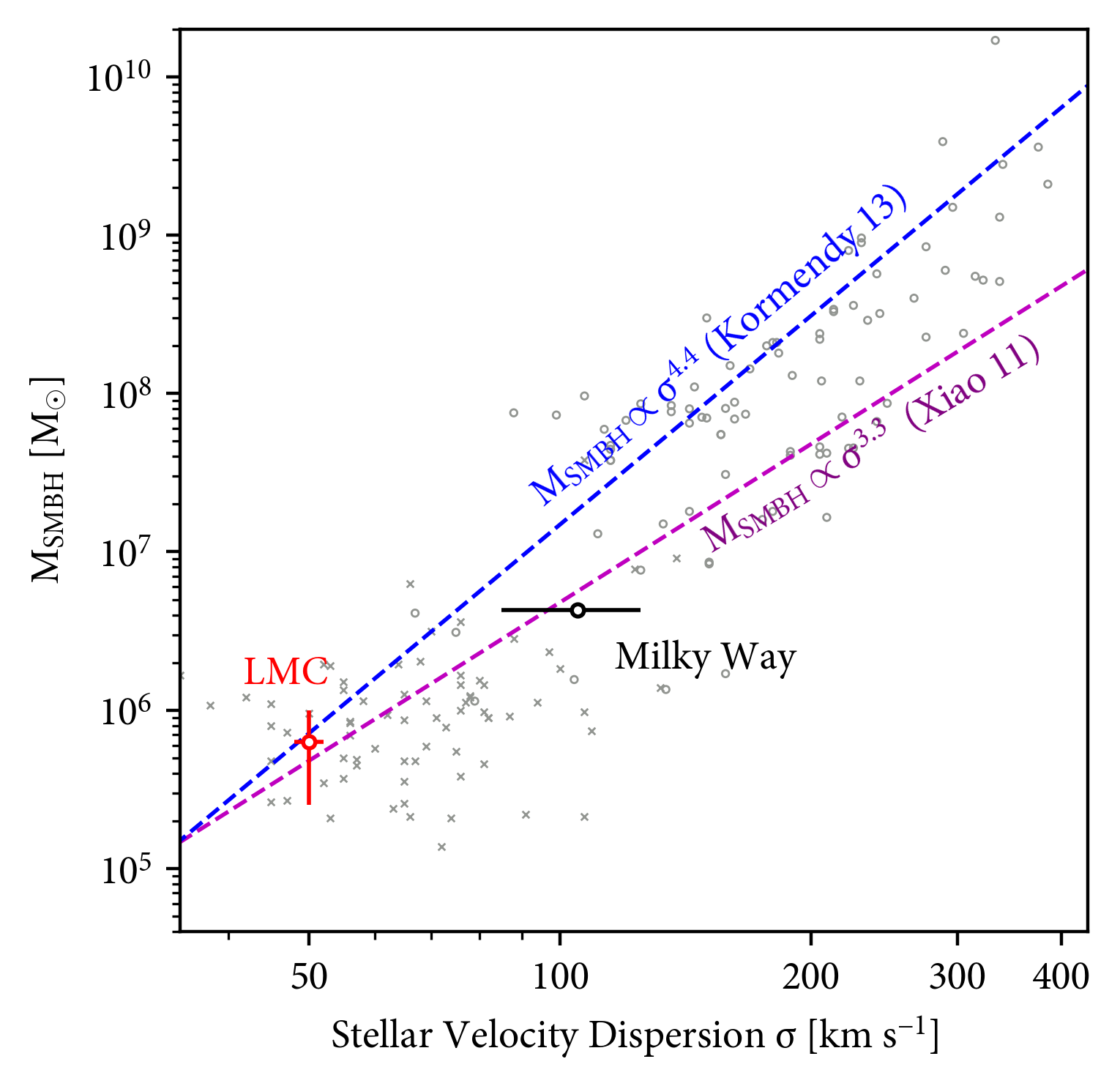}
    \caption{Placing LMC* on the $M-\sigma$ relation. In dashed lines we show two versions of the relation, one fitted to a large sample of galaxies \citep[][light grey circles]{kormendy13} and the other fitted specifically to low-mass SMBHs \citep[][light grey $\times$-marks]{xiao11}. We adopt a velocity dispersion value of $50\pm2\text{ km}\text{ s}^{-1}$ for the LMC \citep{borissova06}. The LMC* mass measured in this study is consistent with both versions of the $M-\sigma$ relation. }
    \label{fig:msigma}
\end{figure}

We also apply the forward model framework to disk runaways originating from the LMC disk and the Galactic disk, ejected at velocities measured for the LMC-origin HVS. We find that (1) if such fast disk runaways exist, they should be readily detectable with the HVS Survey, and (2) the distribution of these stars at present day are significantly more scattered on the sky compared to observations, due to a spread in tangential velocities induced by the LMC disk rotation. We thus conclude that the observed LMC-origin HVSs must primarily be produced by an SMBH in the LMC. We show additional evidence for the existence of LMC* based on HE 0437-5439, which was ejected with too high a velocity to plausibly be explained by anything besides the Hills mechanism (Figure~\ref{fig:hvs3_velocity}). Finally, we produce a grid of simulations across various LMC* masses and binary properties to conduct a parameter search. The key observables from the simulations are the mean ejection velocities from Sgr A* and LMC*, and the count ratio between the two HVS populations. We sample the posterior of the model to constrain the mass of LMC* as $10^{5.8^{+0.2}_{-0.4}} M_{\odot}$ ($\simeq 6\times10^5 M_{\odot}$).

The mass of LMC* measured in this study is significantly larger than what has been previously assumed in the literature (e.g., \citealt{Erkal2019} assume an LMC* mass of $\gtrsim 10^4M_{\odot}$, and \citealt{Gualandris2007} conclude a mass greater than $10^3M_{\odot}$). The discrepancy arises from the fact that we analyze a sample of LMC-origin HVSs instead of a single star. While a lighter black hole---as considered in previous works---can produce one hypervelocity star such as HE 0437-5439, only a supermassive black hole can produce a comparable number of hypervelocity stars to Sgr A*. Meanwhile, direct observational upper limits on the LMC* mass are much higher than any of these values, at $\lesssim 10^{7.1}M_{\odot}$ \citep{boyce17}. 

It is well known that the stellar velocity dispersion and the mass of the supermassive black hole of a galaxy are strongly correlated \cite[e.g.,][]{Magorrian98, ferrarese00, gebhardt00, Gültekin_2009, kormendy13, greene20}. While the LMC does not have a classical bulge, we can use the velocity dispersion of its bar and inner stellar halo ($\sim50\text{ km}\text{ s}^{-1}$ from \citealt{minniti03, borissova06})
to estimate where LMC* would lie on the $M-\sigma$ relation. We show this in Figure \ref{fig:msigma}. In dashed lines we show two versions of the $M-\sigma$ relation: one inferred from a large sample of galaxies \citep{kormendy13}, and the other tailored to low-mass SMBHs \citep{xiao11}. For both relations, $6\times10^{5} M_{\odot}$ falls nearly exactly on the $\sigma=50\text{ km}\text{ s}^{-1}$ line. While these relations have a typical uncertainty of around 0.5 dex (factor of 3 in linear space), it is clear that an LMC* mass of $6\times10^{5} M_{\odot}$  is well within the expected range. Another ``sanity check'' is to simply scale the mass of Sgr A*, $4\times10^6 M_{\odot}$ \citep{ghez08, genzel10} to the stellar mass ratio of the LMC to the Galaxy, which yields $2\times10^5M_{\odot}$. Recalling that Sgr A* falls under the $M-\sigma$ relation by roughly a factor of two \citep{ferrarese00}, one can naively expect an LMC* mass of around $4\times10^5M_{\odot}$, which is within the mass range that we derive. We thus conclude that the LMC* mass derived in this study is fully compatible with the $M-\sigma$ relation.

So far, this study has relied on one model of the LMC orbit from the \citetalias{nico19} simulation. A major uncertainty in the LMC orbit comes from observational uncertainties in the positions, velocities, and masses of the Magellanic Clouds. For example, a 50\% change in the LMC total mass can cause a present-day difference up to $\sim40\text{ km}\text{ s}^{-1}$ in the HVS velocities. Although these variations have a minor impact on the overall HVS population, they can alter the inferred orbits of individual stars traced back to the LMC center. Given these uncertainties, we emphasize that the prediction of the Leo Overdensity shown in Figure \ref{fig:onsky}-\ref{fig:fourpanel} is agnostic to the precise orbit: only an SMBH in the LMC can produce a tight overdensity of hypervelocity stars as observed in the data. Thus, we can instead use the observed LMC-origin HVS to constrain the true orbit of the LMC. The correct orbital history of the LMC-MW system should maximize the overlap of LMC-origin hypervelocity stars with the past locations of the LMC center. We will present the findings of this study in future work.

\begin{acknowledgments}
    We thank Douglas Boubert, Ana Bonaca, and Wenbin Lu for helpful discussions.  This work was made possible through Scialog grant \#SA-LSST-2024-114c from Research Corporation for Science Advancement. LH acknowledges support by the Simons Collaboration on ``Learning the Universe.'' RS is supported by ISF, NSF/BSF, MOS and GIF grants.
\end{acknowledgments}

\bibliography{sample631}{}
\bibliographystyle{aasjournal}

\end{document}